\documentclass[conference]{IEEEtran}
\IEEEoverridecommandlockouts
\usepackage[noadjust]{cite}
\usepackage{amsmath,amssymb,amsfonts}
\usepackage{algorithmic}
\usepackage{graphicx}
\usepackage{textcomp}
\usepackage{xcolor}
\usepackage{ifthen}
\usepackage{enumerate}
\usepackage[ruled,lined]{algorithm2e}
\usepackage{tikz}
\usepackage{subcaption}
\usepackage{pifont}
\usepackage{listings}
\usepackage{lipsum}
\usepackage[hidelinks]{hyperref}
\usepackage{cleveref}
\usepackage{soul}
\usepackage{pifont}
\usepackage{multirow}
\usepackage{booktabs}
\usepackage{makecell}
\usepackage{rotating}

\usepackage[inline]{enumitem}
\usepackage[multiple]{footmisc}

\usepackage{adjustbox}
\usepackage{array}
\newcolumntype{R}[2]{%
    >{\adjustbox{angle=#1,lap=\width-(#2)}\bgroup}%
    l%
    <{\egroup}%
}
\newcommand*\rot{\multicolumn{1}{R{45}{1em}}}

\makeatletter
\newcommand\footnoteref[1]{\protected@xdef\@thefnmark{\ref{#1}}\@footnotemark}
\makeatother

\lstdefinestyle{CStyle}{
    backgroundcolor=\color{backgroundColour},
    commentstyle=\color{mGreen},
    keywordstyle=\color{magenta},
    numberstyle=\tiny\color{mGray},
    stringstyle=\color{mPurple},
    basicstyle=\footnotesize,
    breakatwhitespace=false,
    breaklines=true,
    captionpos=b,
    keepspaces=true,
    numbers=left,
    numbersep=5pt,
    showspaces=false,
    showstringspaces=false,
    showtabs=false,
    tabsize=2,
    language=C
}

\definecolor{mGreen}{rgb}{0,0.6,0}
\definecolor{mGray}{rgb}{0.5,0.5,0.5}
\definecolor{mPurple}{rgb}{0.58,0,0.82}
\definecolor{backgroundColour}{rgb}{0.95,0.95,0.92}
\definecolor{darkgreen}{rgb}{0.0,0.6,0.0}
\definecolor{darkmidnightblue}{rgb}{0.0,0.2,0.4}
\definecolor{crimson}{rgb}{0.86, 0.08, 0.24}
\definecolor{jazzberryjam}{rgb}{0.65, 0.04, 0.37}

\newcommand{\blackball}[1]{\tikz[baseline=(myanchor.base)] \node[circle,fill=.,inner sep=1pt] (myanchor) {\color{-.}\bfseries\footnotesize #1};}

\crefformat{chapter}{\S#2#1#3}
\crefformat{section}{\S#2#1#3}
\crefformat{subsection}{\S#2#1#3}
\crefformat{subsubsection}{\S#2#1#3}

\newcommand{\mynote}[3]{
   \fbox{\bfseries\sffamily\scriptsize#1}
   {\small$\blacktriangleright$\textsf{\emph{\color{#3}{#2}}}$\blacktriangleleft$}}

\newcommand{\tania}[1]{\mynote{tania}{#1}{jazzberryjam}}

\newcommand{\SYS}{\textsc{DIO}}

 \renewcommand{\paragraph}[1]{\vspace{.2\baselineskip}\noindent\textbf{#1.}}

\def\BibTeX{{\rm B\kern-.05em{\sc i\kern-.025em b}\kern-.08em
    T\kern-.1667em\lower.7ex\hbox{E}\kern-.125emX}}

\definecolor{navy}{HTML}{2F729C}

\hypersetup{
    colorlinks=true,
    linkcolor=black,
    urlcolor=navy,
    citecolor=black
}

\begin{document}

\title{Diagnosing applications' I/O behavior through system call observability}

\author{
    \IEEEauthorblockN{Tânia Esteves}
    \IEEEauthorblockA{\textit{INESC TEC \& U. Minho}\\tania.c.araujo@inesctec.pt}
    \and
    \IEEEauthorblockN{Ricardo Macedo}
    \IEEEauthorblockA{\textit{INESC TEC \& U. Minho}\\ricardo.g.macedo@inesctec.pt}
    \and
    \IEEEauthorblockN{Rui Oliveira}
    \IEEEauthorblockA{\textit{INESC TEC \& U. Minho}\\rui.oliveira@inesctec.pt}
    \and
    \IEEEauthorblockN{João Paulo}
    \IEEEauthorblockA{\textit{INESC TEC \& U. Minho}\\jtpaulo@inesctec.pt}
}

\maketitle

\begin{abstract}

We present \SYS, a generic tool for observing inefficient and erroneous I/O interactions between applications and in-kernel storage systems that lead to performance, dependability, and correctness issues. \SYS\space facilitates the analysis and enables near real-time visualization of complex I/O patterns for data-intensive applications generating millions of storage requests. This is achieved by non-intrusively intercepting system calls, enriching collected data with relevant context, and providing timely analysis and visualization for traced events.

We demonstrate its usefulness by analyzing two production-level applications. Results show that \SYS\space enables diagnosing resource contention in multi-threaded I/O that leads to high tail latency and erroneous file accesses that cause data loss.

\end{abstract}


\section{Introduction}
\label{sec:intro}

The performance, correctness and dependability of data-intensive applications (\textit{e.g.}, databases, key-value stores, analytical engines, machine learning frameworks) is highly influenced by the way these interact with in-kernel POSIX storage backends, such as file systems and block devices~\cite{ganesan2017redundancy,roselli2000comparison}.

Due to human error and lack of detailed knowledge on how to efficiently and correctly access the storage backend, developers often implement applications that exhibit: \begin{enumerate*}[label=\textit{\roman*)}]
    \item costly access patterns, such as small-sized I/O requests or random accesses;
    \item I/O contention caused by having concurrent requests accessing shared storage resources; and
    \item erroneous usage of I/O calls, for example, by accessing wrong file offsets.
\end{enumerate*}
These patterns lead to inefficient or incorrect storage I/O accesses, which not only compromise the usefulness of optimizations implemented within each storage backend (\textit{e.g.}, caching, scheduling), but can ultimately degrade end-to-end performance, negatively impact availability, and even cause data loss for applications.

The sheer amount of storage operations generated by these applications, which can range from hundreds to thousands of operations per second, makes their analysis a complex and time-consuming task when done manually. Thus, diagnosis tools that can help users and developers profile more precisely the I/O interaction between applications and corresponding storage backends are crucial for debugging errors, finding performance and dependability issues, and identifying potential optimizations for applications~\cite{daoud2021performance, esteves2021cat, saif2018ioscope}.

The main insight of this paper is that, by combining system call (or \textit{syscall} for short) tracing with a customizable analysis pipeline, one can provide non-intrusive and comprehensive I/O diagnosis for applications using in-kernel POSIX storage systems (\textit{e.g.}, file system, Linux block device).
Doing so requires overcoming the following limitations of existing approaches.

\paragraph{Intrusiveness}
The collection of information about I/O requests is often done through source code instrumentation~\cite{jaeger,zipkin,kim2012iopin, vijayakumar2009scalable}.
This approach is not easily applicable across different applications, as it requires users to manually analyze and instrument distinct and potentially large codebases.

\paragraph{Practicality}
I/O requests can be intercepted non-intrusively with kernel-level tracing technologies. However, the performance penalty imposed on the application by widely-used solutions, such as strace~\cite{strace}, can make this choice unpractical for data-intensive workloads. Namely, it significantly increases the time for tracing requests and, due to the performance slowdown, can hide subtle concurrency issues, such as I/O contention or starvation~\cite{gebai2018survey, esteves2021cat}. This challenge motivated the emergence of technologies such as eBPF~\cite{ebpf} and LTTng~\cite{lttng}, which follow a non-blocking tracing strategy that reduces performance overhead at the cost of potentially discarding I/O events that cannot be processed in a timely fashion.

\paragraph{Lack of analysis pipeline}
While efficient I/O tracing is an important step for profiling applications, by itself is not sufficient, given the large amount of collected events (easily reaching tens of millions) that must be parsed, correlated, and visually represented to provide insightful information (\textit{e.g.}, showcase contention in multi-threaded I/O).
Several solutions only cover the tracing collection step, delegating these other time-consuming tasks to users~\cite{strace,sysdig,akgun2020re,tracee}.

\paragraph{Flexibility}
Solutions offering a complete pipeline for application diagnosis are designed for rigid analysis scenarios, such as detecting unreproducible builds~\cite{ren2019root}, observing file offset access patterns~\cite{saif2018ioscope}, or identifying security issues~\cite{yoo2018longline,persecmon}. Thus, for multi-purpose profiling tasks, one must combine several of these tools and repeat multiple times the tracing, analysis, and visualization of the same application. Ideally, diagnosis tools should provide the flexibility to narrow or broaden both tracing and analysis scopes based on user goals. This would enable, with the same tool, exploring a wider range of performance, correctness, and dependability issues that applications may exhibit, as those identified at \cref{sec:eval}.

This paper proposes \SYS, a generic tool for observing and diagnosing applications' storage I/O, which addresses these challenges with the following contributions.

\paragraph{Non-intrusive, comprehensive, and flexible tracing}
\SYS\space offers a new eBPF-based tracer that intercepts syscalls issued by applications without requiring changes to their source code or instrumentation of binaries. The tracer supports 42 storage-related syscalls and records a comprehensive set of information for each collected operation, including its type, arguments, return value, timestamps, ProcessID (PID), and ThreadID (TID). By offering a flexible design, \SYS\space allows collecting only events of interest, filtering them (in kernel-space) by syscall type, PID, TID, or file paths. This enables narrowing the tracing scope according to users' requirements and minimizing performance overhead over the targeted application.

\paragraph{Enriched analysis}
\SYS\space enriches data gathered for each syscall with additional context available at the kernel (\textit{e.g.,} process name, file type, file offset), which can be used to improve the correlation and analysis of requests (\textit{e.g.}, associating different syscalls to a file path, differentiating operations over regular files or directories). These features enable a richer and wider analysis of incorrect or inefficient I/O patterns.

\paragraph{Asynchronous event handling}
Only syscall interception is done synchronously, while collected events are sent and stored at a remote backend asynchronously. This avoids adding extra latency in the critical path of I/O requests and enables practical analysis of long and data-intensive storage workloads.

\paragraph{Near real-time pipeline}
\SYS\space offers a practical and customizable pipeline so that users can create their own queries, correlation algorithms, and visualization dashboards to analyze collected data. The pipeline follows an in-line approach, meaning that traced information is automatically parsed and forwarded to the analysis and visualization components as soon as it is captured without requiring manual user intervention.

\SYS\space is implemented as an open-source prototype using eBPF~\cite{ebpf}, Elasticsearch~\cite{elasticsearch}, and Kibana~\cite{kibana}, and validated with production-level systems.
Results show that \SYS\space enables the diagnosis of
\begin{enumerate*}[label=\textit{\roman*)}]
    \item erroneous file accesses that cause data loss in Fluent Bit, and
    \item resource contention in multi-threaded I/O that leads to high tail latency for user workloads in RocksDB.
\end{enumerate*}
All artifacts discussed in this paper, including \SYS, workloads, scripts, and the corresponding analysis and visualization outputs are publicly available.\footnote{\url{https://github.com/dsrhaslab/dio}}
\section{The \SYS\space Tool}
\label{sec:design}

\begin{figure}[t]
    \centering
    \includegraphics[width=\columnwidth]{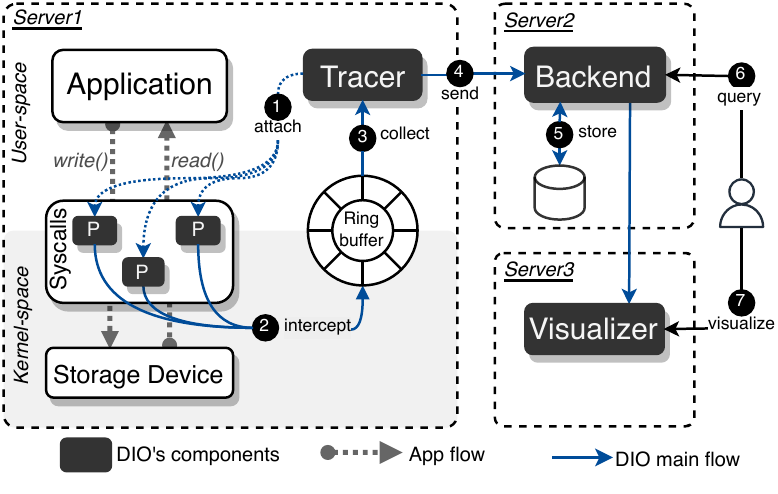}
    \caption{\SYS's design and flow of events.}
    \label{fig:dio-design}
\end{figure}

\SYS\space is a generic tool for observing and diagnosing the I/O interactions between applications and in-kernel POSIX storage systems. Its design is built over the following core principles, which address the challenges discussed in \cref{sec:intro}.

\paragraph{Transparency} \SYS\space relies on the Linux kernel tracing infrastructure, namely tracepoints and kernel probes, to intercept applications' syscalls without requiring any modification to their source code or underlying libraries.

\paragraph{Practical and timely analysis} Traced data is asynchronously sent to a remote analysis pipeline, avoiding adding extra latency on the critical I/O path of applications while enabling users to visualize collected data in near real-time.

\paragraph{Post-mortem analysis} \SYS\space allows storing different tracing executions from the same or different applications and posteriorly analyzing and comparing them.

\paragraph{Flexible and comprehensive tracing}
\SYS\space intercepts different types of storage-related syscalls, covering data (\textit{e.g.}, \texttt{write}), metadata (\textit{e.g.}, \texttt{stat}), extended attributes (\textit{e.g.}, \texttt{getxattr}), and directory management (\textit{e.g.}, \texttt{mknod}) requests. Users can choose to capture only the relevant syscalls for their analysis goals and further filter these based on targeted PIDs, TIDs, or file paths.

\paragraph{Enriching syscall analysis} \SYS\space enriches the information provided directly by each syscall (\textit{i.e.}, type, arguments, return value) with additional context from the kernel, such as the name of the process that originated the request, the type of the file being accessed by it, and its offset.

\paragraph{Data querying and correlation} With \SYS, users can query traced data, apply filters to analyze specific information (\textit{e.g.}, syscalls executed by a specific TID), and correlate different types of data (\textit{e.g.}, associate file descriptors with file paths).

\paragraph{Customized visualization} \SYS\space comprises a visualization component that provides mechanisms for simplifying the exploration of traced data and to build customized visualizations.

\subsection{System overview}

\SYS\space consists of three main components, namely the \textit{tracer}, the \textit{backend}, and the \textit{visualizer}, as depicted in Fig.~\ref{fig:dio-design}. \SYS's analysis pipeline includes the latter two components.

The \textit{tracer} (\cref{sec:design:tracer}) intercepts syscalls from applications, filters them according to the user's configurations (\textit{e.g.}, by TID), and packs their information into events that are asynchronously sent to the \textit{backend} (\blackball{4}).
The \textit{backend} (\cref{sec:design:backend}) persists and indexes events (\blackball{5}), and allows users to query and summarize (\textit{e.g.}, aggregating) stored information (\blackball{6}).

The \textit{visualizer} (\cref{sec:design:viz}) provides near real-time visualization of the traced events by querying the \textit{backend} (\blackball{7}). Users rely on the \textit{visualizer} to ease the process of data exploration and analysis, by selecting specific types of data (\textit{e.g.}, syscall types, arguments) to build different and customized representations.

\subsection{Tracer}
\label{sec:design:tracer}

\begin{table}[t]
\centering
\caption{Syscalls supported by \SYS.}
\label{tab:syscalls}
\resizebox{\columnwidth}{!}{%

\begin{tabular}{cl}

\textbf{Type} &
  \multicolumn{1}{c}{\textbf{Syscall}} \\\cmidrule[0.15em]{1-2}\\[-3mm]

\multicolumn{1}{c}{\multirow{2}{*}{\textit{\textbf{Data}}}} &  \texttt{read, pread64, readv, write, pwrite64} \\
 & \texttt{writev, fsync, fdatasync, readahead} \\[-3mm]\\\cmidrule{2-2}\\[-3mm]

\multicolumn{1}{c}{\multirow{4}{*}{\textit{\textbf{Metadata}}}} & \texttt{creat, open, openat, close} \\
& \texttt{lseek, truncate, ftruncate} \\
 & \texttt{rename, renameat, renameat2} \\
 & \texttt{unlink, unlinkat, readlink, readlinkat} \\
 & \texttt{stat, lstat, fstat, fstatfs, fstatat} \\[-3mm]\\\cmidrule{2-2}\\[-3mm]

\multicolumn{1}{c}{\multirow{3}{*}{\textit{\textbf{Extended}}}}
& \texttt{getxattr, lgetxattr, fgetxattr}\\
& \texttt{setxattr, lsetxattr, fsetxattr}\\
& \texttt{listxattr, llistxattr, flistxattr}\\
& \texttt{removexattr, lremovexattr, fremovexattr} \\[-3mm]\\\cmidrule{2-2}\\[-3mm]

\textit{\textbf{Directory}} & \texttt{mknod, mknodat} \\[-3mm]\\\cmidrule{2-2}\\
\end{tabular}%
}
\end{table}

The \textit{tracer} intercepts syscalls done by applications in a non-intrusive way. To that end, it relies on the eBPF technology~\cite{ebpf}, which allows instrumenting the Linux kernel by executing small programs (\textit{i.e.}, eBPF programs) whenever a given point of interest (\textit{e.g.}, tracepoints, kprobes) is called.

In detail, \SYS's \textit{tracer} comprises a set of eBPF programs that, at the initialization phase (\blackball{1}), are automatically and transparently attached to syscall tracepoints.
Whenever these tracepoints are reached (\textit{i.e.}, a syscall is invoked), the eBPF program gathers the desired information about the request and places it in a per-CPU \textit{ring buffer} (\blackball{2}), which is a contiguous memory area used for exchanging data between kernel (producers) and user-space (consumers) processes.
At user-space, the \textit{tracer} asynchronously fetches information from the \textit{ring buffer} (\blackball{3}), parses it into events (specified in JSON objects), and sends these to the \textit{backend}. To minimize both network and performance overhead, the tracer groups several events into buckets that are sent and indexed in batches at the \textit{backend}.

Table~\ref{tab:syscalls} depicts the syscalls supported by \SYS.
Since instrumenting syscalls can introduce extra processing in the critical path of I/O requests, \SYS\space allows users to filter requests by: \textit{i)} type, \textit{ii)} process or thread ID(s), and \textit{iii)} targeted file or directory path(s). By implementing these filters in the kernel, \SYS\space reduces the amount of information sent to user-space.

\paragraph{Collected information} For each intercepted syscall, \SYS\space collects information related to the: \textit{i)} request (type, arguments, and return value); \textit{ii)} process (PID, TID, and process name) and \textit{iii)} time (\textit{entry} and \textit{exit} timestamps).
While this information alone already provides valuable insights about applications' I/O behavior (\textit{e.g.}, syscalls issued over time, size of I/O requests), correlating it with other types of data further enriches and eases the users' analysis (as further discussed in \cref{sec:eval}).
Therefore, the \textit{tracer} leverages eBPF's access to kernel structures and enriches traced information with:

\begin{itemize}
    \item the \textit{file type} targeted by syscalls, which enables differentiating accesses to regular files, directories, sockets, block/char devices, pipes, symbolic links, and other files.

    \item the \textit{file offset} being accessed by data-related syscalls. This information allows observing file access patterns (\textit{e.g.}, random accesses), even for syscalls that do not provide the file offset as an argument (\textit{e.g.}, \texttt{read}, \texttt{write}).

    \item a \textit{file tag} that labels syscalls handling file descriptors (\textit{e.g.}, \texttt{read}, \texttt{close}) with a tag containing the device number, inode number, and first file access timestamp that uniquely identify the file being accessed.
\end{itemize}

\subsection{Backend}
\label{sec:design:backend}

The \textit{backend} allows persisting, searching, and analyzing data from traced events.
It uses the Elasticsearch~\cite{elasticsearch} distributed engine for storing and processing large volumes of data.
Its flexible document-oriented schema allows indexing events as documents, even if these have potentially different structures (\textit{e.g.}, distinct fields corresponding to syscall arguments). Moreover, it provides an interface for searching, querying, and updating documents, which allows users to develop and integrate customized data correlation algorithms.

\paragraph{File path correlation algorithm}
We have implemented a custom algorithm to enable the correlation of syscalls with specific accessed file paths. Using Elasticsearch's data querying and updating features, the file tags (\textit{i.e.}, unique identifiers generated by the \textit{tracer} component) associated with syscalls are translated into the actual file paths being accessed at the storage backend (\textit{e.g.}, \texttt{/tmp/fileA}).


\subsection{Visualizer}
\label{sec:design:viz}

The \textit{visualizer} provides an automated approach towards exploring (\textit{e.g.}, query and filter events) and visually depicting (\textit{e.g.}, through tables, histograms, time-series graphs) the analysis findings.
This component uses Kibana~\cite{kibana}, the data visualization dashboard software for Elasticsearch, which is often used for log and time-series analytics and application monitoring. Kibana also allows building custom visualizations, thus being aligned with the design principles of \SYS.

\subsection{Implementation}
The tracer is implemented in $\approx$8K LoC, divided into two parts: \textit{i)} the eBPF programs that run in kernel-space and \textit{ii)} the user-space code including the remaining tracer's logic.

The eBPF programs are implemented in C and are responsible for collecting and filtering relevant storage I/O events.
The user-space code is implemented in Go (v17.4) and is responsible for enabling the desired I/O tracepoints (attaching eBPF programs), specifying the user-defined filters applied at each tracepoint, gathering and parsing the information collected at kernel-space, and sending it to the \textit{backend} component.
This is done using the BPF Compiler Collection (BCC) framework through the \textit{gobpf} lib (v0.2.0), which provides an abstraction for creating, attaching, and interacting with eBPF programs. For communication with Elasticsearch, we use the \textit{go-elasticsearch} (v7.13.1) module, taking advantage of the bulk indexing API for sending multiple events simultaneously.

The \textit{backend} and \textit{visualizer} components use Elasticsearch (v8.5.2) and Kibana (v8.5.2), respectively. The file path correlation algorithm can be automatically executed by the \textit{tracer} or on-demand by users.

\subsection{Configuration and Usage}

The installation and configuration of \SYS\space are performed in two phases: \textit{i)} the setup and initialization of the analysis pipeline and \textit{ii)} the configuration and execution of the \textit{tracer}.

\paragraph{Analysis pipeline}
Although all \SYS's components can be deployed in the same machine, to avoid negatively impacting the performance of the targeted application (\textit{e.g.}, additional resource consumption), the analysis pipeline can be installed on a separate server(s) (Fig.~\ref{fig:dio-design}).
Further, as the \textit{tracer} component labels each tracing execution with a unique session name, one can deploy \SYS\space as a service, setting up the analysis pipeline on dedicated servers and allowing multiple executions of \SYS's \textit{tracer} on different machines and by distinct users.

The deployment and configuration of the analysis pipeline comprise its software installation (\textit{i.e.}, Elasticsearch and Kibana) and importing its predefined dashboards. As soon as tracing data arrives at the pipeline, users can access Kibana's web page and visualize \SYS's dashboards, apply analysis filters, and edit or create new visualizations and dashboards.

\paragraph{Tracer}
Once the analysis pipeline is deployed, users can use \SYS's \textit{tracer} to collect information. The \textit{tracer} executes along with the targeted application, stopping once its main and child processes finish or upon explicit users' instruction.

By default, \SYS's \textit{tracer} enables tracepoints for the full set of supported syscalls. However, users can specify a list of syscalls to observe, and the \textit{tracer} will only activate tracepoints for those operations. Also, users may specify a list of files/directories to observe, instructing the \textit{tracer} to only record events that target them. All these configurations, along with the analysis pipeline's parameters (\textit{e.g.}, Elasticsearch URL), can be set through a configuration file.

\section{Evaluation}
\label{sec:eval}

Our evaluation showcases how \SYS\space can be used to ease the process of observing/confirming known issues and validating their fixes. To this end, we analyzed two production-level applications: Fluent Bit and RocksDB. Results show that \SYS\space is a practical tool for validating the root causes of correctness (\cref{sec:eval:fb}) and performance (\cref{sec:eval:rocksdb}) issues, without instrumenting large codebases.

With the exception of Fig.~\ref{fig:rocksdb_latency}, all the remaining figures in this section were generated by \SYS\space (with minimal modifications for readability).
The full set of \SYS's visual representations is available at \url{https://github.com/dsrhaslab/dio}.

\subsection{Experimental Setup}

Our testbed comprises three servers running \textit{Ubuntu 20.04 LTS} with kernel \textit{5.4.0}. The server running the application and \SYS's \textit{tracer} is equipped with
a 4-core Intel Core i3-7100, 16 GiB of memory, a 250 GiB NVMe SSD (which hosts the datasets), and a 512 GiB SATA SSD (used for logging).
\SYS's \textit{backend} and \textit{visualization} components run on two separated servers, both equipped with a 6-core Intel i5-9500, 16 GiB of memory, and a 250 GiB NVMe SSD.

\subsection{Identifying erroneous actions that lead to data loss}
\label{sec:eval:fb}

\begin{figure*}[t!]
    \centering
    \begin{subfigure}[t]{0.5\textwidth}
        \centering
        \includegraphics[width=\columnwidth]{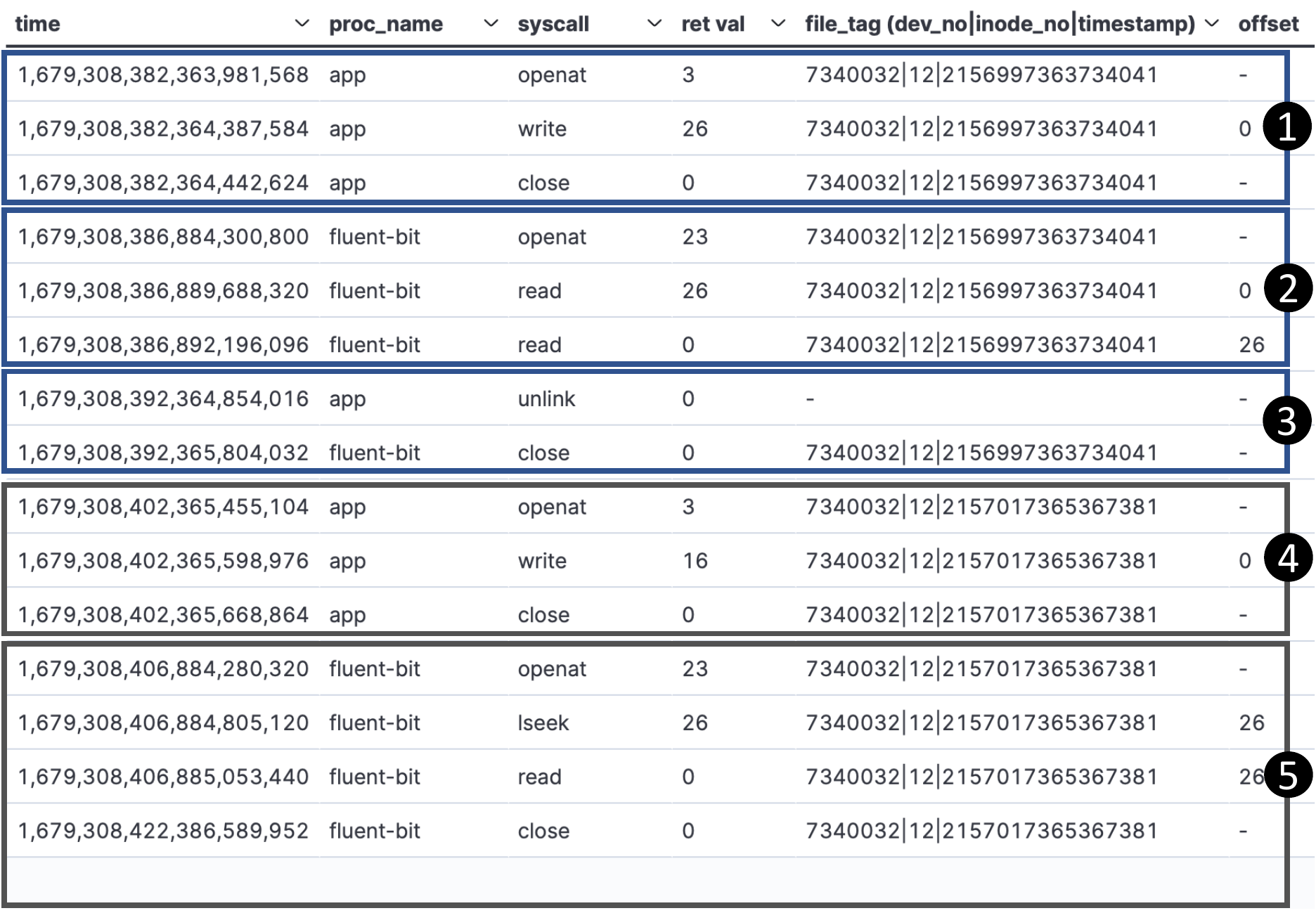}
        \caption{Fluent Bit (v1.4.0) erroneous access pattern.}
        \label{fig:fluent-bit:v1}
    \end{subfigure}%
    ~
    \begin{subfigure}[t]{0.5\textwidth}
        \centering
        \includegraphics[width=\columnwidth]{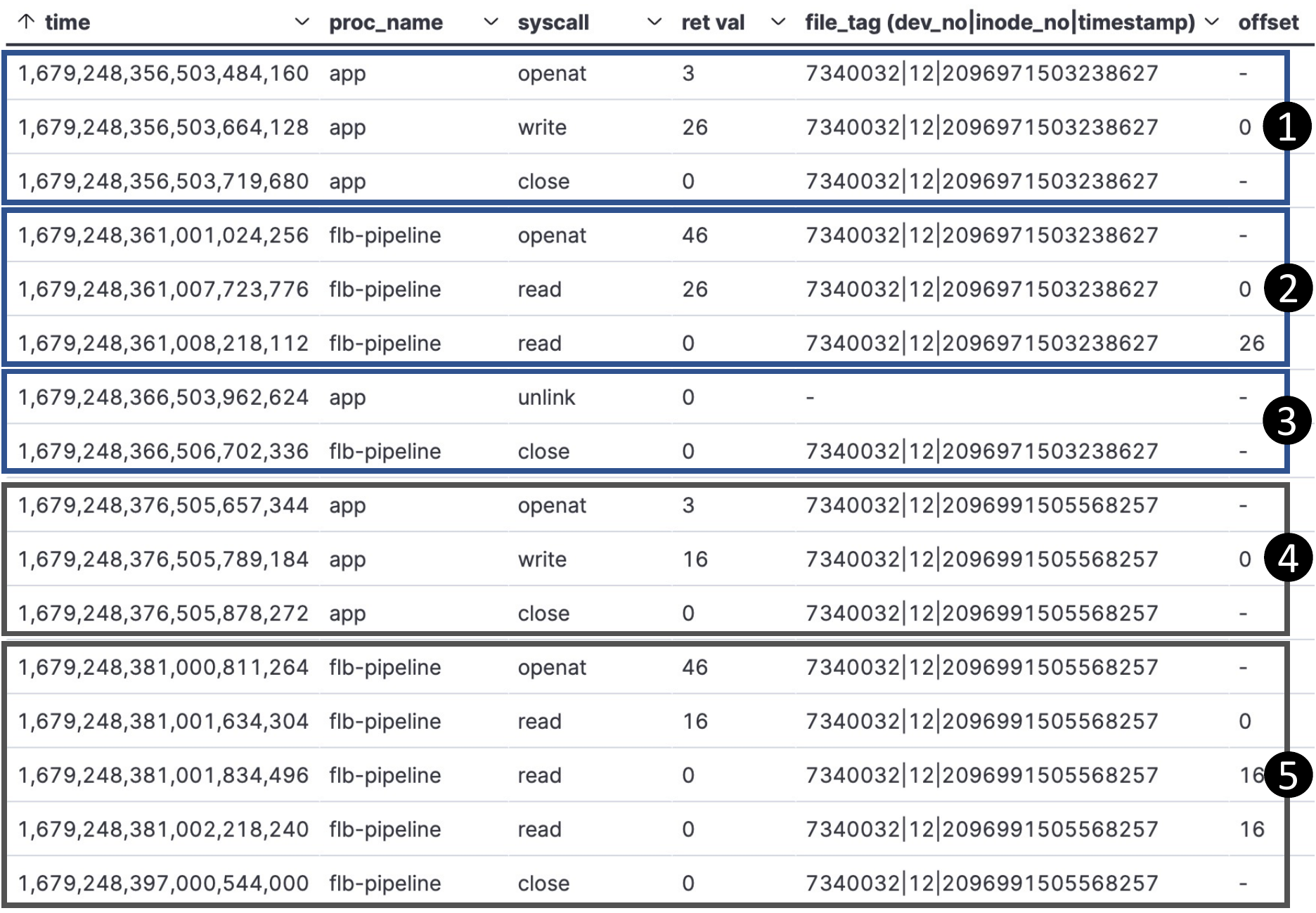}
        \caption{Fluent Bit (v2.0.5) correct access pattern.}
        \label{fig:fluent-bit:v2}
    \end{subfigure}
    \caption{Fluent Bit erroneous access pattern leading to data loss.}
    \label{fig:fluent-bit}
\end{figure*}

\SYS\space can assist developers and users in diagnosing the correctness of their applications.
We demonstrate this by showing erroneous I/O access patterns that result in data loss.

For this use case, we consider Fluent Bit~\cite{fluentbit} (v1.4.0), a high-performance logging and metrics processor and forwarder.
Existing issues\footnote{\label{fn:fb1}\url{https://github.com/fluent/fluent-bit/issues/1875}}\textsuperscript{,}\footnote{\url{https://github.com/fluent/fluent-bit/issues/4895}} report that data is lost when using the \texttt{tail} input plugin, which is used to fetch new content being added to log files.
Thus, we implemented a client program that simulates the generation of log files to be processed by Fluent Bit and mimics the I/O behavior reported in Issue \#1875\footnoteref{fn:fb1}.
\SYS\space was used to simultaneously trace and analyze the client program and Fluent Bit by filtering the syscalls belonging to the set of processes of these applications.

Fig.~\ref{fig:fluent-bit:v1} shows a tabular visualization generated by \SYS. The client program (\textit{app}) starts by creating the \texttt{app.log} file, writing 26 bytes starting from offset 0, and closing the file (\blackball{1}).
Then, Fluent Bit (\textit{fluent-bit}) detects content modification at the file, opens it, and reads 26 bytes from offset 0, which means that \textit{fluent-bit} processes the full content previously written by \textit{app} (\blackball{2}).
Later, \textit{app} removes the file with the \texttt{unlink} syscall, and \textit{fluent-bit} closes the corresponding file descriptor (\blackball{3}). At the operating system level, this means that the inode number associated with this file (\texttt{12}) is now unused and will later be attributed to a new file.
However, a possible scenario is this inode number being mapped to a newly created file with the same name.
This happens when \textit{app} creates a new file with the same name as the previous one (\texttt{app.log}) and writes 16 bytes to it (\blackball{4}).
The incorrect behavior reported at the issue, and observable with \SYS, happens when \textit{fluent-bit} opens the new log file for reading its content, but instead of reading from offset 0, as expected, it starts reading at offset 26 (\blackball{5}). By starting at the wrong offset, the \texttt{read} syscall returns zero bytes, and the 16 bytes written by \textit{app} are lost.

To understand the reason for this behavior, we examined Fluent Bit's code responsible for reading new content entries in log files. Before reading a file, Fluent Bit updates the file position to the number of bytes already processed. This value is kept on a database for each tracked file, identified by its name plus inode number.
Erroneously, database entries are not deleted when files are removed from the file system. Therefore, and going back to our running example, since the same file name (\texttt{app.log}) and inode number (\texttt{12}) are attributed to the newly created file, \textit{fluent-bit} erroneously assumes that the first 26 bytes of the latter log file were already processed.

To validate the correction of this erroneous access pattern, we used \SYS\space to analyze a more recent version of Fluent Bit (v2.0.5), where fixes were applied to avoid this data loss issue. Fig.~\ref{fig:fluent-bit:v2} shows a similar tabular visualization for the fixed version. While the two versions present similar behavior (same file accesses for \blackball{1}-\blackball{4}), the difference relies on the file offset being accessed by Fluent Bit (\textit{flb-pipeline}) when reading from a new file (\blackball{5}). This time, Fluent Bit starts reading from the beginning of the file (offset 0), being able to read the new 16 bytes written by \textit{app}.

This example shows that \SYS\space helps users diagnose incorrect I/O behavior from applications and find the root cause for dependability issues such as data loss.
Further, while this example only showcases a small amount of lost data, it can be significantly higher when dealing with larger log files.
Moreover, this use case also exemplifies how \SYS\space helps validate the corrections applied to the implementation of applications.

\subsection{Finding the root cause of performance anomalies}
\label{sec:eval:rocksdb}

We now demonstrate how \SYS\space can also ease the process of diagnosing performance issues by identifying the root cause for high tail latency at client requests issued to RocksDB, an embedded key-value store (KVS)~\cite{rocksdb}. 

\begin{figure}[!t]
    \centering
    \includegraphics[width=\columnwidth]{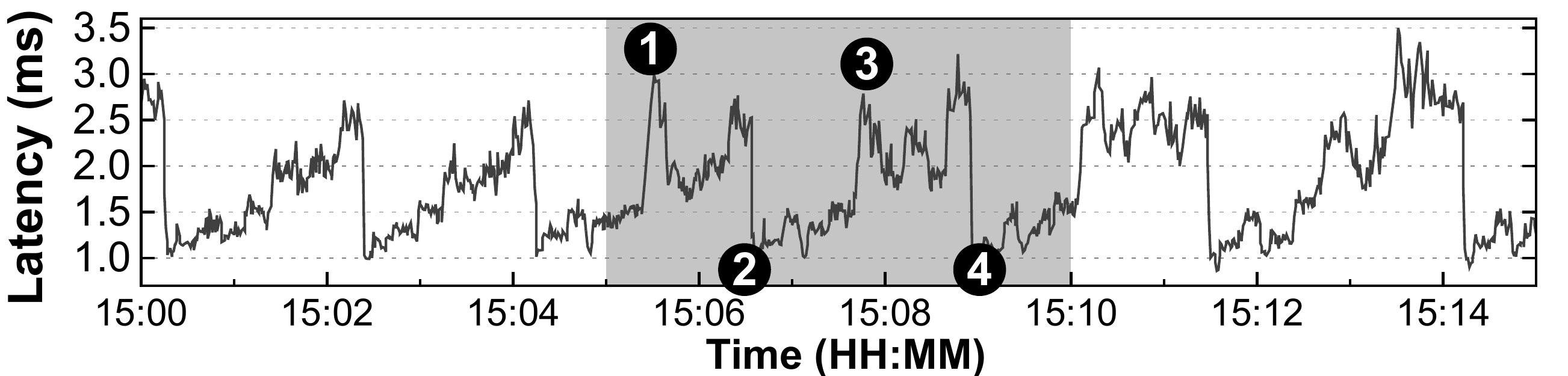}
    \caption{99$^{th}$ percentile latency for RocksDB client operations.}
    \label{fig:rocksdb_latency}
\end{figure}

This phenomenon was first observed in SILK~\cite{balmau2019silk} and, therefore, we followed the same testing methodology to reproduce it. We used the \textit{db\_bench} benchmark~\cite{dbbench} configured with 8 client threads performing a mixture of read-write requests in a closed loop (YCSB A~\cite{ycsba}). RocksDB was configured with 8 background threads, namely 1 for flushes and 7 for compactions.
Fig.~\ref{fig:rocksdb_latency} reports a sample of a 5-hour long execution and depicts the 99$^{th}$ percentile latency experienced by clients. 
Throughout this sample, clients observe several latency spikes that range between 1.5~ms to 3.5~ms.

\begin{figure*}[!t]
    \centering
    \includegraphics[width=\textwidth]{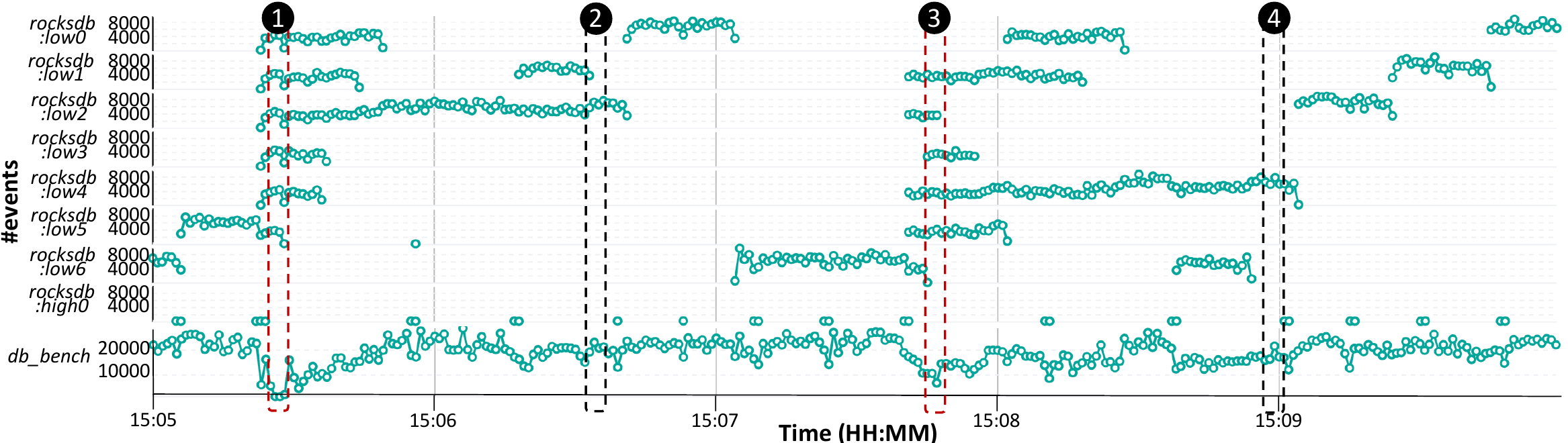}
    \caption{Syscalls issued by RocksDB over time, aggregated by thread name. \texttt{db\_bench} includes the 8 client threads, \texttt{rocksdb:low[0-6]} refers to each compaction thread, and \texttt{rocksdb:high0} refers to the flush thread.
    }
    \label{fig:rocksdb-tids}
\end{figure*}

Finding the root cause for this performance penalty through RocksDB codebase instrumentation would require inspecting more than 440K LoC and adding debugging code to several core components. 
Alternatively, with \SYS, one can easily trace, analyze, and visualize RocksDB execution, as depicted in Fig.~\ref{fig:rocksdb-tids}.
Since the workload is data-oriented, we configured \SYS's \textit{tracer} to capture exclusively \texttt{open}, \texttt{read}, \texttt{write}, and \texttt{close} syscalls.
Client threads are represented as \texttt{db\_bench}, while \texttt{rocksdb:high0} respects to the flushing thread, and the remainder (\texttt{rocksdb:lowX}) to compaction threads.

By observing the syscalls submitted over time by different RocksDB threads, one can identify performance contention. 
Namely, as shown by the highlighted red boxes, when multiple compaction threads submit I/O requests, the number of syscalls of \texttt{db\_bench} threads decreases, causing an immediate tail latency spike perceived by clients, as depicted in Figs.~\ref{fig:rocksdb_latency} and~\ref{fig:rocksdb-tids} (in intervals \blackball{1} and \blackball{3}, at least 5 compaction threads submit requests).
When fewer compaction threads perform I/O, the performance of \texttt{db\_bench} threads improves both in terms of tail latency and throughput (in intervals \blackball{2} and \blackball{4}, only 1 to 2 compaction threads are performing I/O).

If one complements the previous observation with knowledge of how Log Structured Merge-tree (LSM) KVSs work, the problem becomes clear: RocksDB uses foreground threads to process client requests (\texttt{db\_bench} threads), which are enqueued and served in FIFO order. 
In parallel, background threads serve internal operations, namely flushes (\texttt{rocksdb:high0}) and compactions (\texttt{rocksdb:lowX}). 
Flushes ensure that in-memory key-value pairs are sequentially written to the first level of the persistent LSM tree ($L_0$), and these can only proceed when there is enough space at $L_0$. 
Compactions are held in a FIFO queue, waiting to be executed by a dedicated thread pool. 
Except for low-level compactions ($L_0$$\rightarrow$$L_1$), these can be made in parallel. 
A common problem of compactions, however, is the interference between I/O workflows, generating latency spikes for client requests. 
Specifically, latency spikes occur when client threads cannot proceed because $L_0$$\rightarrow$$L_1$ compactions and flushes are slow or on hold, which happens, for instance, when several threads compete for shared disk bandwidth (creating contention).
This is precisely the phenomenon identified in SILK, which can negatively impact the response time and even the availability of KVSs and services that use them~\cite{TaleScale:2013:Dean,TalesOfTheTail:2014:Li}, and that can be observed with \SYS\space without any code instrumentation. 

\subsection{Performance impact and I/O events handling}

\begin{table}[t]
\centering
\caption{Average execution time and standard deviation for 3 independent runs of RocksDB.}
\label{tab:perf}
\resizebox{\columnwidth}{!}{%
\begin{tabular}{ccccc}
 & \textit{vanilla} & \textit{sysdig} & \textit{DIO} & \textit{strace} \\\cmidrule[0.15em]{2-5}\\[-3mm]
\textbf{Average execution time} & 03h48m & 03h56m & 05h12m & 06h30m \\\\[-3mm]
\textbf{Standard deviation} & $\pm$1.2m & $\pm$1.7m & $\pm$2.4m & $\pm$4.6m \\\cmidrule{2-5}\\[-3mm]
\textbf{Overhead} &  & 1.04x & 1.37 & 1.71x \\\cmidrule{2-5}\\[-3mm]
\end{tabular}
}

\end{table}

To understand the performance impact induced by \SYS\space when intercepting I/O syscalls, we selected the RocksDB use case, which includes a benchmark, and measured the average execution time of three independent runs.

\paragraph{Performance analysis} Table~\ref{tab:perf} compares the \textit{vanilla} deployment (\textit{i.e.}, without tracing its execution) with \SYS, \textit{strace}~\cite{strace} -- a widely used syscall tracer, and \textit{Sysdig}~\cite{sysdig} -- a state-of-the-art eBPF-based syscall tracer.

Executing the \textit{vanilla} setup requires approximately 3 hours and 48 minutes. Compared to \textit{vanilla}, \textit{Sysdig} imposes the smallest overhead (1.04x), using 8 minutes more to execute, while \textit{strace} has the highest overhead (1.71x), lasting for more 2 hours and 42 minutes. \SYS\space increases performance overhead by 1.37x, needing extra 1 hour and 16 minutes when compared to \textit{Sysdig}, but saving up to 1 hour and 18 minutes regarding \textit{strace}.

The difference between \textit{strace} and the two eBPF-based tracers (i.e., \textit{sysdig} and \SYS) can be explained by the underlying tracing technology. The trap mechanism used to intercept syscalls and the context switching done by \textit{strace} impose considerable overhead over the targeted application~\cite{gebai2018survey}.

Regarding the eBPF-based tracers, \textit{Sysdig} presents smaller performance overhead than \SYS, but it also reports less information to users. Namely, while \SYS\space is unable to report file paths for up to 5\% of the collected events (due to discarded events), \textit{Sysdig} is unable to report these for 45\% of the events. As expected, making more information available to the user induces a higher performance penalty in the applications' execution time.

\paragraph{I/O events handling}
As discussed in \cref{sec:design}, \SYS\space uses a fixed-sized ring buffer to collect information at user-space, which was configured with 256 MiB per CPU core for these experiments. When this buffer is full (\textit{i.e.}, if kernel processes are producing I/O events to the ring buffer at a faster pace than the user-space processes can consume them), new I/O events being intercepted at the kernel level are discarded.

For the RocksDB experiments, given its intensive I/O behavior, 3.5\% of the issued syscalls ($\approx$19M of 549M) were discarded at the ring buffer and, therefore, not stored at \SYS's \textit{backend}.

\subsection{Summary}

The previous use cases demonstrate \SYS's capabilities for diagnosing distinct I/O patterns, which enables users to observe applications' I/O behaviors, confirm known issues, and validate the correction of their fixes.
Moreover, our integrated tracing and analysis pipeline allows users to observe these I/O patterns without resorting to code instrumentation or needing to manually combine multiple tools.

Our preliminary experimental results show that \SYS\space can collect, parse, and forward to the analysis pipeline all the required tracing information while imposing reduced performance overhead.
Further, despite the discarded I/O events in RocksDB, we show that \SYS\space is still able to pinpoint resource contention and help diagnose its root cause. Moreover, unlike in \textit{strace} and \textit{sysdig}, \SYS's traced information is made available for visualization as soon as it is intercepted and transmitted to the \textit{backend} component.

\section{Related Work}

\paragraph{I/O tracing} Storage I/O diagnosis is often done by capturing applications' requests in user-space through source code instrumentation~\cite{jaeger,zipkin,kim2012iopin, vijayakumar2009scalable}; through middleware libraries~\cite{snyder2016modular,naas2021eziotracer} that are restricted to specific sets of applications (\textit{e.g.}, \textit{LD\_PRELOAD} only works with dynamic libraries);
or at lower kernel layers~\cite{saif2018ioscope,persecmon,naas2021eziotracer}, such as the Virtual File System, where optimizations like I/O merging make it impossible to observe the exact requests submitted by applications.

To intercept I/O operations non-intrusively and closer to the requests made by applications, other solutions rely on the syscall interface. As shown in Table~\ref{tab:rw}, these explore distinct tracing technologies, including ptrace (\cite{strace,ren2019root}), eBPF (\cite{sysdig, tracee, esteves2021cat}), LTTng (\cite{akgun2020re, daoud2021performance, kohyarnejadfard2021framework}), and auditd (\cite{yoo2018longline}), which allow gathering information related with the \textit{entry} and \textit{exit} points of syscalls, including their arguments, return value, timestamps, PIDs, \textit{etc}. Similar to \SYS, some tools enrich traced data with additional information such as the \textit{process name} (\cite{sysdig, tracee, esteves2021cat, yoo2018longline}), which is useful for observing the I/O patterns at \cref{sec:eval:fb}, and \cref{sec:eval:rocksdb}. However, \SYS\space is the only tool that collects \textit{file offsets}, which are crucial for diagnosing the use case presented in \cref{sec:eval:fb}.

Only CaT~\cite{esteves2021cat}, Tracee~\cite{tracee}, and \SYS~\space aggregate the information contained at the \textit{entry} and \textit{exit} points of each syscall into a single event, thus simplifying its posterior analysis. This is done at kernel-space to reduce the data transferred to user-space. Further, these are the only tools, along with \textit{strace}~\cite{strace} and \textit{Sysdig}~\cite{sysdig}, that support filtering at the tracing phase.

\paragraph{Integrated analysis pipeline} Several solutions only cover the tracing step, leaving the integration with analysis pipelines to be done by users~\cite{strace, sysdig, akgun2020re, tracee}. Other tools provide modules for automating the analysis of traced data but follow an offline approach, where this data needs to be stored first and, only later, it is parsed and provided as input to the analysis pipeline~\cite{ren2019root, esteves2021cat, daoud2021performance, kohyarnejadfard2021framework}. Only \SYS\space and Longline~\cite{yoo2018longline} automatically parse and forward traced events to the analysis pipeline by following an inline (near real-time) approach. 

\paragraph{Syscall analysis} Some of the existing tools support analysis modules specialized for their concrete use cases (\textit{e.g.}, causality~\cite{ren2019root, esteves2021cat}, security analysis~\cite{yoo2018longline}), which only consider specific information collected from traces (\textit{e.g.}, syscall types). Therefore, these do not provide the flexibility to implement custom analysis algorithms nor enable users to access and explore other information contained in the collected I/O traces. On the other hand, solutions similar to \SYS\space that support customizable analysis fail to capture relevant information to diagnose the use cases discussed in this paper~\cite{daoud2021performance, kohyarnejadfard2021framework}.

\SYS\space provides users access to the complete set of captured information (\textit{e.g.}, syscall type, arguments, offsets), allowing them to build new algorithms over the data fields that are more relevant to their analysis goals.

\paragraph{Syscall visualization} 
\SYS\space offers predefined representations that automatically summarize and allow the visualization of the I/O patterns discussed in the paper. Moreover, our tool enables users to create new visualizations commonly supported by other diagnosis solutions (\textit{e.g.}, tables, pie charts, histograms, heatmaps, time series)~\cite{daoud2021performance, kohyarnejadfard2021framework, yoo2018longline}.

\renewcommand*\rot[2]{\multicolumn{1}{R{#1}{#2}}}

\begin{table}[t]
\centering
\caption{Comparison between \SYS\space and other solutions in terms of: captured tracing information, filtering capabilities, tracing and analysis integration (O-offline, I-inline), analysis customization, and predefined visualization support. While some tools are able to trace (T) the information required for the paper's use-cases, only \SYS\space provides users with the analysis (A) capabilities to diagnose them.}
\label{tab:rw}
\resizebox{\columnwidth}{!}{%
\setlength\tabcolsep{3pt}
\begin{tabular}{cccccccccccc}

\multicolumn{2}{l}{}
& \rot{60}{1em}{\scriptsize \textit{Strace}\cite{strace}}
& \rot{60}{1em}{\scriptsize \textit{Sysdig}\cite{sysdig}}
& \rot{60}{1em}{\scriptsize \textit{Re-Animator}\cite{akgun2020re}}
& \rot{60}{1em}{\scriptsize \textit{RepTrace}\cite{ren2019root}}
& \rot{60}{1em}{\scriptsize \textit{Tracee}\cite{tracee}}
& \rot{60}{1em}{\scriptsize \textit{CAT}\cite{esteves2021cat}}
& \rot{60}{1em}{\scriptsize \cite{daoud2021performance}}
& \rot{60}{1em}{\scriptsize \cite{kohyarnejadfard2021framework}}
& \rot{60}{1em}{\scriptsize \textit{LongLine}\cite{yoo2018longline}}
& \rot{60}{1em}{\scriptsize \textit{DIO}}  \\\cmidrule[0.15em]{2-12}\\[-1.5mm]

\multicolumn{1}{c}{\multirow{5}{*}{\rotatebox{90}{\begin{tabular}[c]{@{}c@{}}\textit{\textbf{Tracing}}\end{tabular}}}} & \multicolumn{1}{c}{Syscall info} & \ding{51} & \ding{51} & \ding{51} & \ding{51} & \ding{51} & \ding{51} & \ding{51} & \ding{51} & \ding{51} & \multicolumn{1}{c}{\ding{51}} \\[-0.5mm]

\multicolumn{1}{c}{} & \multicolumn{1}{c}{\textit{f\_offset}} & - & - & - & - & - & - & - & - & - & \multicolumn{1}{c}{\ding{51}} \\[-0.5mm]

\multicolumn{1}{c}{} & \multicolumn{1}{c}{\textit{f\_type}} & - & \ding{51} & - & - & - & - & - & - & - & \multicolumn{1}{c}{\ding{51}} \\[-0.5mm]

\multicolumn{1}{c}{} & \multicolumn{1}{c}{\textit{proc\_name}} & - & \ding{51} & - & - & \ding{51} & \ding{51} & - & - & \ding{51} & \multicolumn{1}{c}{\ding{51}} \\[-0.5mm]

\multicolumn{1}{c}{} & \multicolumn{1}{c}{Filters} & \ding{51} & \ding{51} & - & - & \ding{51} & \ding{51} & - & - & - & \multicolumn{1}{c}{\ding{51}} \\[-1.5mm]\\\cline{2-12}\\[-1.5mm]

\multicolumn{1}{c}{\multirow{3}{*}{\rotatebox{90}{\begin{tabular}[c]{@{}c@{}}\textit{\textbf{Analysis}}\\ \textit{\textbf{pipeline}}\end{tabular}}}} & \multicolumn{1}{c}{Integrated} & - & - & - & O & - & O & O & O & I & \multicolumn{1}{c}{I} \\

\multicolumn{1}{c}{} &  \multicolumn{1}{c}{\begin{tabular}[c]{@{}c@{}}Customizable\end{tabular}} & - & - & - & - & - & - & \ding{51} & \ding{51} & - & \multicolumn{1}{c}{\ding{51}} \\

\multicolumn{1}{c}{} & \multicolumn{1}{c}{\begin{tabular}[c]{@{}c@{}}Predefined vis.\end{tabular}} & - & - & - & - & - & \ding{51} & \ding{51} & \ding{51} & \ding{51} & \multicolumn{1}{c}{\ding{51}} \\[-1.5mm]\\\cline{2-12}\\[-1mm]

\multicolumn{1}{c}{\multirow{2}{*}{\rotatebox{90}{\begin{tabular}[c]{@{}c@{}}\textit{\textbf{Use}}\\ \textit{\textbf{cases}}\end{tabular}}}} & \multicolumn{1}{c}{\cref{sec:eval:rocksdb}} & - & T & - & - & T & T & - & - & T & \multicolumn{1}{c}{TA} \\

\multicolumn{1}{c}{} & \multicolumn{1}{c}{\cref{sec:eval:fb}} & - & - & - & - & - & - & - & - & - & \multicolumn{1}{c}{TA} \\[-1.5mm]\\\cline{2-12}

\end{tabular}%
}
\end{table}
\section{Future directions}
\label{sec:future_directions}

By intercepting applications' I/O syscalls in a non-intrusive way and automatically parsing and forwarding the collected data to an analysis pipeline, \SYS\space saves users the time needed to understand the applications' source code, instrument the relevant parts, and manually parse the resulting data.

While the current prototype already provides relevant and summarized information about the targeted application, it would be interesting to further simplify the analysis process for users with, for instance, new automated correlation algorithms.
Therefore, as a future direction, we intend to explore the Elasticsearch query API and build a collection of correlation algorithms that can, for instance, quickly identify the inefficient behaviors observed in the aforementioned applications.

Moreover, we plan to expand \SYS's scope to showcase its capabilities for exploring other applications, even when users are unfamiliar with these, while potentially uncovering new I/O patterns and unidentified issues regarding the performance, dependability, correctness, and security of such applications.

Finally, we aim to analyze the performance overhead imposed by \SYS\space over targeted applications in more detail and study new optimizations that minimize this penalty and reduce the number of I/O events discarded at the tracing phase.
\section{Conclusion}

This paper presents \SYS, a generic tool for observing and diagnosing I/O interactions between applications and in-kernel POSIX storage systems. Through a pipeline that automates the process of tracing, filtering, correlating, and visualizing millions of syscalls, and by enriching the information provided by these with additional context, \SYS\space helps users observing I/O issues while reducing the search space for finding their root cause when, for instance, source code inspection is required.  

Our experiments with two widely-used systems show that \SYS\space provides key information for observing erroneous I/O access patterns that lead to data loss, and identifying resource contention in multi-threaded I/O that leads to high tail latency.

\bibliographystyle{IEEEtran}
\bibliography{dio}

\end{document}